\documentclass[twocolumn,nofootinbib]{revtex4}
\usepackage{graphicx}
\usepackage{latexsym}

\def\be{\begin{equation}}
\def\ee{\end{equation}}
\def\bea{\begin{eqnarray}}
\def\eea{\end{eqnarray}}
\newcommand{\DE}{{\mathrm{de}}}
\newcommand{\DM}{{\mathrm{dm}}}

\begin{document}

\title{Determining Cosmological Parameters with Latest Observational Data}

\author{Jun-Qing Xia${}^{a,c}$}
\author{Hong Li${}^{b,c}$}
\author{Gong-Bo Zhao${}^{d}$}
\author{Xinmin Zhang${}^{a,c}$}

\affiliation{${}^a$Institute of High Energy Physics, Chinese
Academy of Science, P.O.Box 918-4, Beijing 100049, P.R.China}

\affiliation{${}^b$Department of Astronomy, School of Physics,
Peking University, Beijing 100871, P.R.China}

\affiliation{${}^c$Theoretical Physics Center for Science
Facilities (TPCSF), Chinese Academy of Science, P.R.China}

\affiliation{${}^d$Department of Physics, Simon Fraser University,
Burnaby, BC V5A 1S6, Canada}

%\date{\today}

\begin{abstract}

In this paper, we combine the latest observational data, including
the WMAP five-year data (WMAP5), BOOMERanG, CBI, VSA, ACBAR, as
well as the Baryon Acoustic Oscillations (BAO) and Type Ia
Supernoave (SN) ``Union" compilation (307 sample), and use the
Markov Chain Monte Carlo method to determine the cosmological
parameters, such as the equation-of-state (EoS) of dark energy,
the curvature of universe, the total neutrino mass and the
parameters associated with the power spectrum of primordial
fluctuations. Our results show that the $\Lambda$CDM model remains
a good fit to the current data. In a flat universe, we obtain the
tight limit on the constant EoS of dark energy as,
$w=-0.977\pm0.056$ ($1~\sigma$). For the dynamical dark energy
models with time evolving EoS parameterized as $w_{\DE}(a) = w_0 +
w_1 (1-a)$, we find that the best-fit values are $w_0=-1.08$ and
$w_1=0.368$, implying the preference of Quintom model whose EoS
gets across the cosmological constant boundary during evolution.
For the curvature of universe $\Omega_k$, our results give
$-0.012<\Omega_k<0.009$ ($95\%$ C.L.) when fixing $w_{\DE}=-1$.
When considering the dynamics of dark energy, the flat universe is
still a good fit to the current data, $-0.015<\Omega_k<0.018$
($95\%$ C.L.). Regarding the neutrino mass limit, we obtain the
upper limits, $\sum m_{\nu}<0.533$ eV ($95\%$ C.L.) within the
framework of the flat $\Lambda$CDM model. When adding the SDSS
Lyman-$\alpha$ forest power spectrum data, the constraint on $\sum
m_{\nu}$ can be significantly improved, $\sum m_{\nu}<0.161$ eV
($95\%$ C.L.). However, these limits can be weakened by a factor
of $2$ in the framework of dynamical dark energy models, due to
the obvious degeneracy between neutrino mass and the EoS of dark
energy model. Assuming that the primordial fluctuations are
adiabatic with a power law spectrum, within the $\Lambda$CDM
model, we find that the upper limit on the ratio of the tensor to
scalar is $r<0.200$ ($95\%$ C.L.) and the inflationary models with
the slope $n_s\geq1$ are excluded at more than $2~\sigma$
confidence level. However, in the framework of dynamical dark
energy models, the allowed region in the parameter space of
($n_s$,$r$) is enlarged significantly. Finally, we find no strong
evidence for the large running of the spectral index,
$\alpha_s=-0.019\pm0.017$ ($1~\sigma$) for the $\Lambda$CDM model
and $\alpha_s=-0.023\pm0.019$ ($1~\sigma$) for the dynamical dark
energy model, respectively.

\end{abstract}

%\pacs{98.80.Es; 98.80.Cq}

\maketitle

%Introduction==========================================================

\section{Introduction}
\label{Int}

With the accumulation of observational data from cosmic microwave
background measurements (CMB), large scale structure surveys (LSS)
and supernovae observations and the improvements of the data
quality, the cosmological observations play a crucial role in our
understanding of the universe, and also in constraining the
cosmological parameters, such as the EoS of dark energy models,
the curvature of universe, the total neutrino masses and those
associated with the running of the spectral index and
gravitational waves. In our previous work \cite{Xiaplanck}, we
have used the Markov Chain Monte Carlo (MCMC) method to constrain
cosmological models from the astronomical observational data,
including the WMAP three-year data (WMAP3)
\cite{WMAP3GF,WMAP3Other}, small-scale CMB data, LSS data
\cite{SDSS,2df} and SNIa ESSENCE sample \cite{Essence}. We found
that the cosmological constant is consistent with the data,
however the dynamical dark energy models are still allowed and
interestingly the model with its EoS getting across $w = -1$ the
{\it Quintom} model \cite{Quintom} is the best fit model. And we
found no strong significant evidence for the non-flat universe and
massive neutrino. Within the $\Lambda$CDM model, the
scale-invariant spectrum and the spectra with $n_s>1$ are
disfavored by more than $2~\sigma$ confidence level. Due to the
degeneracy between the EoS of dark energy and tensor fluctuation,
those inflationary models excluded within the $\Lambda$CDM model,
will revive in the framework of dynamical dark energy models.
Furthermore, we did not find any significant evidence for the
tensor fluctuations and the large running of the spectral index.

Given the precision of current observations, these results are not
conclusive for the time being. Recently, the WMAP experiment has
published its five-year data of temperature and polarization power
spectra \cite{WMAP5GF1,WMAP5GF2,WMAP5Other}. The Arcminute
Cosmology Bolometer Array (ACBAR) experiment has also published
its new CMB temperature power spectrum \cite{ACBAR}. These new CMB
data can strengthen the constraints on the cosmological
parameters, especially for the inflationary models
\cite{KinneyWMAP5,WMAP5GF1,WMAP5GF2}. Furthermore, the Supernova
Cosmology Project has made an unified analysis of the world's
supernovae datasets and presented a new compilation ``Union" (307
sample) \cite{Union} which includes the recent samples of SNIa
from SNLS and ESSENCE Survey, as well as some older datasets,
\emph{etc}. In the literature \cite{Union,LinderUnion}, this Union
compilation combining with the shift parameter derived from CMB
and the BAO information has been used to constrain cosmological
models. However, in these studies the CMB information considered
is just the shift parameter instead of the full CMB data, which
will lose some information of CMB and lead to a biased result
\cite{Compare}. Thus, in this paper, we revisit the issue on the
determination of these cosmological parameters and update our
previous results with the latest observational data.

Our paper is organized as follows: In Section II we describe the
method and the latest observational datasets we used; Section III
contains our main global fitting results on the cosmological
parameters and the last section is the summary.

%Method and Current Observations=======================================

\section{Method and Data}
\label{Method}

In our study, we perform a global analysis using the publicly
available MCMC package CosmoMC\footnote{Available at:
http://cosmologist.info/cosmomc/.} \cite{CosmoMC}. We assume the
purely adiabatic initial conditions. Our most general parameter
space is:
\begin{equation}
\label{parameter} {\bf P} \equiv (\omega_{b}, \omega_{c},
\Omega_k, \Theta_{s}, \tau, w_{0}, w_{1}, f_{\nu}, n_{s}, A_{s},
\alpha_s, r)~,
\end{equation}
where $\omega_{b}\equiv\Omega_{b}h^{2}$ and
$\omega_{c}\equiv\Omega_{c}h^{2}$, in which $\Omega_{b}$ and
$\Omega_{c}$ are the physical baryon and cold dark matter
densities relative to the critical density, $\Omega_k$ is the
spatial curvature and satisfies
$\Omega_k+\Omega_m+\Omega_{\DE}=1$, $\Theta_{s}$ is the ratio
(multiplied by 100) of the sound horizon to the angular diameter
distance at decoupling, $\tau$ is the optical depth to
re-ionization, $f_{\nu}$ is the dark matter neutrino fraction at
present, namely,
\begin{equation}
f_{\nu}\equiv\frac{\rho_{\nu}}{\rho_{\DM}}=\frac{\Sigma
m_{\nu}}{93.105~\mathrm{eV}~\Omega_ch^2}~.
\end{equation}
The primordial scalar power spectrum $\mathcal{P}_{\chi}(k)$ is
parameterized as \cite{Ps}:
\begin{eqnarray}
\ln\mathcal{P}_{\chi}(k)=\ln
A_s(k_{s0})&+&(n_s(k_{s0})-1)\ln\left(\frac{k}{k_{s0}}\right)\nonumber\\
&+&\frac{\alpha_s}{2}\left(\ln\left(\frac{k}{k_{s0}}\right)\right)^2~,
\end{eqnarray}
where $A_s$ is defined as the amplitude of initial power spectrum,
$n_s$ measures the spectral index, $\alpha_{s}$ is the running of
the scalar spectral index and $r$ is the tensor to scalar ratio of
the primordial spectrum. For the pivot scale we set
$k_{s0}=0.05$Mpc$^{-1}$. Moreover, $w_0$ and $w_1$ are the
parameters of dark energy EoS, which is given by
\cite{Linderpara}:
\begin{equation}
\label{EOS} w_\DE(a) = w_{0} + w_{1}(1-a)~,
\end{equation}
where $a=1/(1+z)$ is the scale factor and $w_{1}=-dw/da$
characterizes the ``running" of EoS (RunW henceforth). The
$\Lambda$CDM model has $w_0=-1$ and $w_1=0$. For the dark energy
model with a constant EoS, $w_1=0$ (WCDM henceforth). When using
the global fitting strategy to constrain the cosmological
parameters, it is crucial to include dark energy perturbations
\cite{WMAP3GF,LewisPert,XiaPert}. In this paper we use the method
provided in Refs.\cite{XiaPert,ZhaoPert} to treat the dark energy
perturbations consistently in the whole parameter space in the
numerical calculations.

In the computation of CMB we have included the WMAP5 temperature
and polarization power spectra with the routine for computing the
likelihood supplied by the WMAP team\footnote{Available at the
LAMBDA website: http://lambda.gsfc.nasa.gov/.}. We also include
some small-scale CMB measurements, such as BOOMERanG
\cite{BOOMERanG}, CBI \cite{CBI}, VSA \cite{VSA} and the newly
released ACBAR data \cite{ACBAR}. Besides the CMB information, we
also combine the distance measurements from BAO and SNIa. For the
BAO information, we use the gaussian priors on the distance
ratios, $r_s/D_v(z)=0.1980\pm0.0058$ at $z=0.2$ and
$r_s/D_v(z)=0.1094\pm0.0033$ at $z=0.35$, with a correlation
coefficient of $0.39$, measured from the BAO in the distribution
of galaxies \cite{BAO}. In the calculation of the likelihood from
SNIa we have marginalized over the nuisance parameter
\cite{SNMethod}. The supernova data we use are the recently
released ``Union" compilation (307 sample) \cite{Union}. In order
to improve the constraint on the total neutrino mass, we include
the Lyman-$\alpha$ forest power spectrum from Sloan Digital Sky
Survey (SDSS) \cite{Lya}, however also keep its unclear
systematics in mind \cite{WMAP5GF1,FogliWMAP5}. Furthermore, we
make use of the Hubble Space Telescope (HST) measurement of the
Hubble parameter $H_{0}\equiv 100$h~km~s$^{-1}$~Mpc$^{-1}$ by a
Gaussian likelihood function centered around $h=0.72$ and with a
standard deviation $\sigma=0.08$ \cite{HST}.

%Results===============================================================

\section{Numerical Results}
%Results(Dark Energy)==================================================

In this section we present our global fitting results of the
cosmological parameters determined from the latest observational
data and focus on the dark energy parameters, curvature of
universe, neutrino mass and the inflationary parameters,
respectively.

\begin{figure}[htbp]
\begin{center}
\includegraphics[scale=0.4]{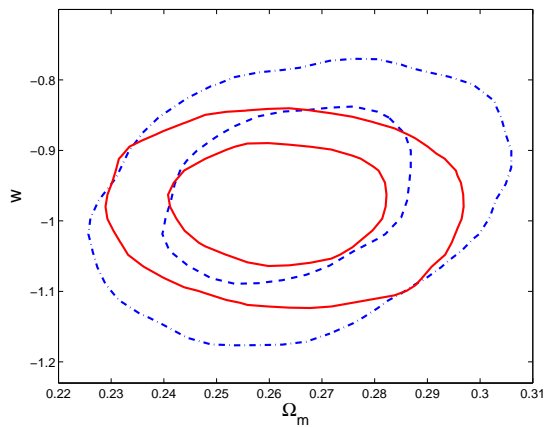}
\caption{Constraints on the constant EoS of dark energy, $w$, and
the present matter density, $\Omega_m$, from the latest
observations, assuming a flat universe. The blue dash-dot lines
and red solid lines are obtained with and without the systematic
uncertainties of Union compilation, respectively.\label{fig1}}
\end{center}
\end{figure}

\begin{figure}[htbp]
\begin{center}
\includegraphics[scale=0.4]{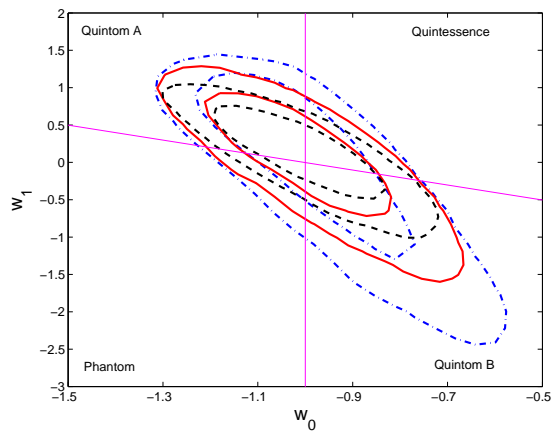}
\caption{Constraints on the dark energy EoS parameters $w_0$ and
$w_1$ from the current observations, CMB+BAO+SN. The red solid
lines and the blue dash-dot lines are obtained for the flat and
non-flat universe, respectively. And the black dashed lines are
obtained when (incorrectly) neglecting dark energy perturbations.
The magenta solid lines stand for $w_0=-1$ and $w_0+w_1=-1$. In
this numerical calculation the systematic uncertainties of Union
compilation is not considered. \label{fig2}}
\end{center}
\end{figure}

\subsection{Equation of State of Dark Energy}
\label{DEEoS}

\begin{table*}{\footnotesize
TABLE I. Constraints on the dark energy EoS and some background
parameters from the latest observations. Here we have shown the
mean and the best fit values, which are obtained from the cases
with and without the systematic uncertainties of Union
compilation, respectively.
\begin{center}

\begin{tabular}{|c|c|c|c|c|c|c|c|c|c|c|c|}

\hline

\multicolumn{2}{|c|}{Parameter}&\multicolumn{2}{c|}{$w_0$}&\multicolumn{2}{c|}{$w_1$}&\multicolumn{2}{c|}{$\Omega_\DE$}&\multicolumn{2}{c|}{$H_0$}\\

\cline{3-10}

\multicolumn{2}{|c|}{}& with sys & w/o sys & with sys & w/o sys &
with sys & w/o sys & with sys & w/o sys \\

\hline

$\Lambda$CDM & BestFit & $-1$ & $-1$ & $0$ & $0$ & $0.735$ & $0.741$ & $71.0$ & $71.6$\\

\cline{2-10}

$\Omega_k=0$& Mean & $-1$ & $-1$ & $0$ & $0$ & $0.738\pm0.015$ & $0.738\pm0.014$ & $71.4\pm1.4$ & $71.4\pm1.3$\\

\hline

WCDM & BestFit & $-0.978$ & $-0.955$ & $0$ & $0$ & $0.738$ & $0.735$ & $71.4$ & $70.6$\\

\cline{2-10}

$\Omega_k=0$& Mean & $-0.965\pm0.080$ & $-0.977\pm0.056$ & $0$ & $0$ & $0.736\pm0.016$ & $0.737\pm0.014$ & $70.8\pm1.9$ & $71.1\pm1.4$\\

\hline

RunW & BestFit & $-1.09$ & $-1.08$ & $0.533$ & $0.368$ & $0.735$ & $0.738$ & $70.4$ & $71.1$\\

\cline{2-10}

$\Omega_k=0$& Mean & $-0.946\pm0.194$ & $-0.993\pm0.128$ & $-0.133\pm0.749$ & $0.030\pm0.582$ & $0.734\pm0.017$ & $0.737\pm0.014$ & $70.7\pm1.9$& $70.9\pm1.5$ \\

\hline

RunW & BestFit & $-$ & $-1.11$ & $-$ & $0.475$ & $-$ & $0.739$ & $-$ & $72.4$\\

\cline{2-10}

$\Omega_k\neq0$& Mean & $-$ & $-0.976\pm0.148$ & $-$ & $-0.071\pm0.848$ & $-$ & $0.736\pm0.014$ & $-$& $70.9\pm1.9$ \\

\hline

RunW & BestFit & $-$ & $-1.04$ & $-$ & $0.290$ & $-$ & $0.742$ & $-$ & $71.3$\\

\cline{2-10}

w/o Pert.& Mean & $-$ & $-1.00\pm0.114$ & $-$ & $0.103\pm0.413$ & $-$ & $0.736\pm0.012$ & $-$& $70.8\pm1.5$ \\

\hline
\end{tabular}
\end{center}}
\end{table*}

In Table I we list the constraints on the dark energy parameters
as well as the Hubble constant in different dark energy models.

Assuming the flat universe, firstly we explore the constraints on
the constant EoS of dark energy, $w$ ($w\equiv
w_0$,~$w_1\equiv0$), from the latest observational data. In
Fig.\ref{fig1} we show the constraints on $w$ and the present dark
matter density, $\Omega_m$. This result shows that the combination
of Union compilation (with systematic uncertainties not included)
and other observational data yield a strong constraint on the
constant EoS of dark energy, $w=-0.977\pm0.056$ ($1~\sigma$). Our
result is similar to the limit from WMAP5 \cite{WMAP5GF1},
$w=-0.972^{+0.061}_{-0.060}$ ($1~\sigma$) and physically it
indicates that $w=-1$ is consistent with the data. Furthermore,
some of the quintessence models get strongly constrained, for
example, the tracker quintessence model which predicts $w\sim-0.7$
\cite{Tracking} will be excluded by more than $5~\sigma$. However
we notice that the systematic uncertainties of Union compilation
will affect strongly the error estimation of dark energy
parameters \cite{Union}. If taking the systematic uncertainties
into account, we find the limit on the constant EoS of dark energy
is $w=-0.965\pm0.080$ ($1~\sigma$) and the error bar is
significantly enlarged.

For the time evolving EoS, $w_\DE(a)=w_0+w_1(1-a)$, in
Fig.\ref{fig2} we illustrate the constraints on the dark energy
parameters $w_0$ and $w_1$. For the flat universe, from the latest
observational data we find that the best fit model is the Quintom
dark energy model, $w_0=-1.08$ and $w_1=0.368$, whose $w(z)$ can
cross the cosmological constant boundary during the evolution. But
then the variance of $w_0$ and $w_1$ are still large, namely, the
$95\%$ constraints on $w_0$ and $w_1$ are $-1.22<w_0<-0.721$ and
$-1.33<w_1<0.947$. This result implies that the dynamical dark
energy models are not excluded and the current data cannot
distinguish different dark energy models decisively. The
$\Lambda$CDM model, however is still a good fit right now.

Our results are consistent with the WMAP5 group \cite{WMAP5GF1},
while the upper limit on $w_0$ and lower limit on $w_1$ are
slightly weaker than theirs. This difference is mainly from the
supernovae datasets used. The supernovae dataset we use in this
paper is the new Union compilation with homogeneous analysis of
the present world's supernovae data. But in Ref.\cite{WMAP5GF1}
they use three supernovae datasets, SNLS \cite{SNLS}, HST
\cite{Riess2006} and ESSENCE \cite{Essence}. For each of them,
they marginalize over the absolute magnitude separately, and
simply add these three pieces to get the total $\chi^2$.

Because the parametrization of EoS of the dark energy used in this
paper is assumed to extend to an arbitrary high redshift, it is
important to check if the energy density of dark energy component
is negligible compared with the radiation density at the epoch of
the Big Bang Nucleosynthesis (BBN), $z\sim10^9$. As shown by the
red solid lines in Fig.\ref{fig2}, the dynamical dark energy
models allowed by the current data within the $95\%$ confidence
level safely satisfy the limits of $w_0+w_1<0$ and BBN
\cite{BBN,Wright} to avoid the dark energy domination in the early
universe.

Furthermore, the Null Energy Condition (NEC) should also be
satisfied for the EoS of universe $w_u$ \cite{NEC}:
\begin{equation}
w_u(a)\equiv\frac{\sum w_i(a)\rho_i(a)}{\sum
\rho_i(a)}\geq-1~,\label{NEC}
\end{equation}
where $w_i$ and $\rho_i$ are the EoS and energy density for
component $i$ in the universe. Violation of NEC will lead to the
breakdown of causality in general relativity and the violation of
the second law of thermodynamics \cite{NECVio}. This requirement
from $w_u(a)\geq-1$ will constrain the EoS parameters of dark
energy models \cite{QiuNEClimit}.

Firstly, we consider the WCDM dark energy model. From Table I, the
current observational data give the present energy density of dark
energy $\Omega_\DE\simeq0.74$. If we just assume that the EoS of
dark energy is constant from the early time of universe to
present, using the above Eq.(\ref{NEC}), we obtain that the NEC
limit requires the EoS of dark energy $w_\DE\gtrsim-1.35$
straightforwardly \cite{Creminelli:2006xe}. Fortunately, the
current constraints on $w_\DE$ with and without the systematic
uncertainties of ``Union" compilation satisfy this limit safely.
However, if we extend the period of validity of constant EoS of
dark energy into the future where the dark energy component will
dominate the universe entirely (namely $\Omega_\DE\approx1$), the
NEC requires $w_\DE\gtrsim-1$. Consequently, the phantom dark
energy models ($w_\DE<-1$) will be excluded and the fate of
universe with Big Rip \cite{BigRip} could not happen.

\begin{figure}[htbp]
\begin{center}
\includegraphics[scale=0.4]{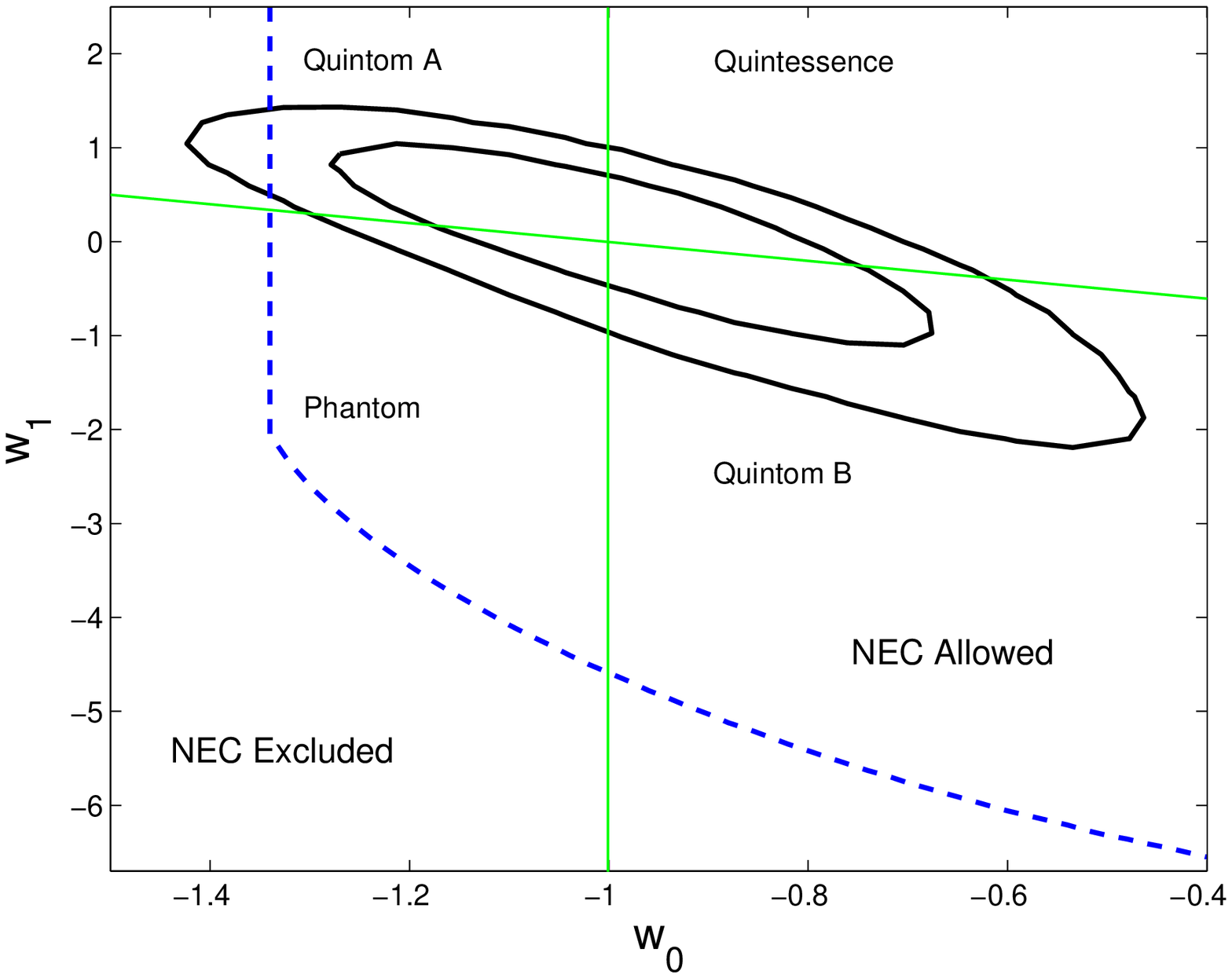}
\includegraphics[scale=0.4]{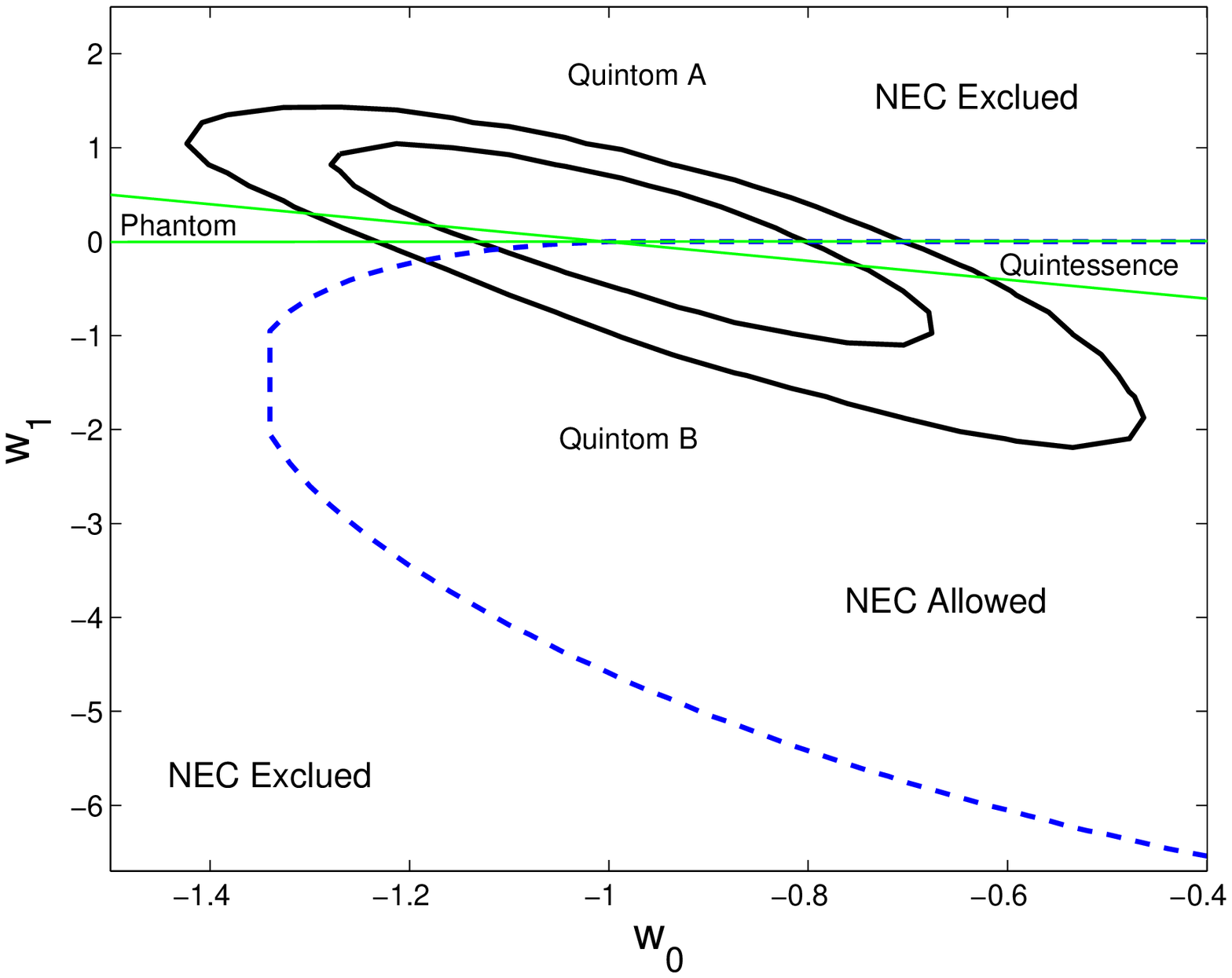}
\caption{Constraints on the dark energy EoS parameters $w_0$ and
$w_1$ from the current observations (black solid lines) and the Null
Energy Condition (blue dashed lines). In the upper panel, the
parametrization of EoS in Eq.(\ref{EOS}) is assumed to be a
description of dark energy for the past until now. In the lower
panel, we assume that the parametrized EoS are also valid for the
far future. \label{figNEC}}
\end{center}
\end{figure}

For the RunW dark energy model, we illustrate the constraints on
the dark energy parameters $w_0$ and $w_1$ from the current
observations (black solid lines) and the NEC (blue dashed lines)
in Fig.\ref{figNEC}, respectively. In the upper panel, the
parameterized EoS $w_\DE=w_0+w_1(1-a)$ is assumed to be valid
until now, not for the future, {\it i.e.} $0\leq{a}\leq1$. The
reasons for doing this are i) experimentally there are no
constraints on the dark energy models for the future universe; ii)
theoretically it is always possible to choose a different type of
parametrization of dark energy EoS for the future universe, such
as $w_\DE=w_0\exp(1-a)$ for $a>1$, which matches to
$w_\DE=w_0+w_1(1-a)$ at $a=1$ and satisfies the NEC
\cite{QiuNEClimit}. In fact, if the current EoS of dark energy
$w_0<-1$, this scenario allows the transition from $w_\DE<-1$ to
$w_\DE>-1$, consequently avoids the violation of NEC. One can see
from the Fig.\ref{figNEC}, in this case, the NEC limit does
improve the constraints on dark energy parameters from the current
observations. For example, the current observational data permit
$w_0<-1.35$ at $2~\sigma$ confidence level, which however violates
the NEC limit.

In the lower panel of Fig.\ref{figNEC}, the parametrization
$w_\DE=w_0+w_1(1-a)$ is assumed to be a valid description of dark
energy model at any time from the far past to the far future. One
can see from this figure, the NEC puts a stronger constraint on
the EoS parameters of dark energy models than the observational
data. For this case, the regions of quintessence and phantom dark
energy models will be shrunk significantly in the ($w_0$,$w_1$)
space, and the NEC excludes the dark energy models corresponding
to the regions labelled by the phantom and Quintom A with
$w_\DE<-1$ for $a\rightarrow\infty$. Consequently, the models
satisfying the NEC include the quintessence and some of the
Quintom B dark energy models, which can be seen in the lower panel
of Fig.\ref{figNEC}.

Finally we discuss the degeneracy between the dark energy and
curvature. As we know, the EoS of dark energy is degenerated with
the curvature of universe $\Omega_k$ \cite{ZhaoOmk,Bassett}. If we
do not include the prior that the universe is flat, from Table I
and the blue dash-dot lines in Fig.\ref{fig2}, we can see that the
constraints on $w_0$ and $w_1$ are weakened significantly and the
two-dimensional distribution extends more towards the Quintom B
region. But the main conclusions are unchanged, $w_0=-1.11$ and
$w_1=0.475$, namely, the Quintom dark energy model is still mildly
favored by the current observational data. Moreover, in order to
show the importance of dark energy perturbations in the global
analysis, we do the calculation by incorrectly neglecting the dark
energy perturbations. Illustrated as the black dashed lines in
Fig.\ref{fig2}, one can see that the constraints on the dark
energy parameters become tighter immediately, similar to our
previous results \cite{XiaPert}. This study shows that how biased
the result will be, once the dark energy perturbations are
incorrectly neglected in the analysis
\cite{WMAP3GF,XiaPert,Compare}.

\begin{table}
TABLE II. Constraints on cosmological parameters $n_s$,
$\alpha_s$, $r$, $\Omega_k$ and $\sum m_{\nu}$ from the current
observations. We have shown the mean $1,2~\sigma$ errors. For the
weakly constrained parameters we quote the $95\%$ upper limits
instead.
\begin{center}

\begin{tabular}{|c|c|c|}

\hline

Parameters & $\Lambda$CDM & RunW \\

\hline

$100\times\Omega_k$&$-0.081^{+0.545+1.025}_{-0.524-1.161}$&$0.098^{+0.881+1.655}_{-0.881-1.605}$\\

\hline

$\sum m_{\nu}$&$<0.533$ ($95\%$)&$<0.974$ ($95\%$)\\

\hline

$\sum m_{\nu}$ (w/ Ly$\alpha$) &$<0.161$ ($95\%$)&$<0.252$ ($95\%$)\\

\hline

$n_s$&$0.961^{+0.012+0.024}_{-0.012-0.023}$&$0.964^{+0.013+0.027}_{-0.013-0.025}$\\

\hline

$\alpha_s$&$-0.019^{+0.017+0.032}_{-0.017-0.030}$&$-0.023^{+0.019+0.039}_{-0.019-0.037}$\\

\hline

$r$&$<0.200$ ($95\%$)&$<0.268$ ($95\%$)\\

\hline
\end{tabular}
\end{center}
\end{table}

%Results(Others,Curvature)=============================================

\subsection{Curvature of Universe} \label{Omk}

\begin{figure}[t]
\begin{center}
\includegraphics[scale=0.5]{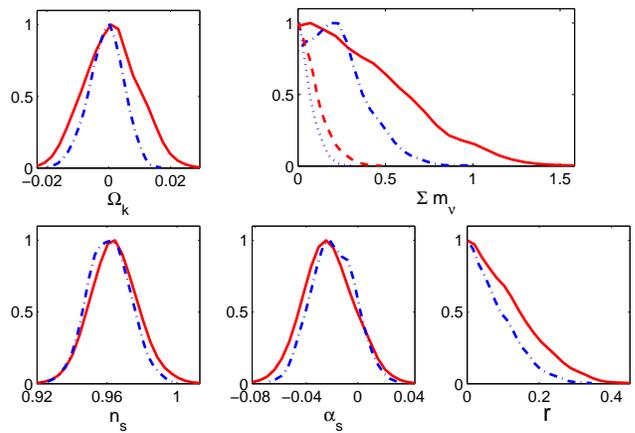}
\caption{$1D$ current constraints on the inflationary parameters
$n_s$, $\alpha_s$ and $r$, as well as the curvature $\Omega_k$ and
the total neutrino mass $\sum m_{\nu}$, in different dark energy
models: $\Lambda$CDM model (blue dash-dot lines) and RunW model
(red solid lines). For $\sum m_{\nu}$, we also show the limits
combined with the SDSS Lyman-$\alpha$ forest power spectrum. Blue
dotted lines and red dashed lines denote the $\Lambda$CDM and RunW
models, respectively.\label{fig3}}
\end{center}
\end{figure}

The measurements on the position of first acoustic peak of CMB
temperature power spectrum have been used to determine the
curvature of universe $\Omega_k$. However, due to the well-known
degeneracy between $\Omega_m$ and $\Omega_k$, we have to add other
cosmological data, such as the large scale structure and
supernovae data, to break this degeneracy and improve the
constraint. Within the $\Lambda$CDM model, from Table II and
Fig.\ref{fig3}, we can see that our universe is very close to
flatness, namely, the $95\%$ limit is $-0.012<\Omega_k<0.009$,
which is consistent with the prediction of inflation paradigm.

As we mentioned before, dark energy parameters and $\Omega_k$ are
correlated via the cosmological distance information. In the
framework of dynamical dark energy models, the constraint on
$\Omega_k$ should be relaxed. Based on the calculations, we can
see that the combination of observational data implies
$-0.015<\Omega_k<0.018$ at $2~\sigma$ confidence level.

%Results(Others,Neutrino)==============================================

\subsection{Neutrino Mass} \label{Mnu}

Detecting the neutrino mass is one of the challenges of modern
physics. Currently the neutrino oscillation experiments, such as
atmospheric neutrinos experiments \cite{Atmospheric} and solar
neutrinos experiments \cite{Solar}, have confirmed that the
neutrinos are massive, but give no hint on their absolute mass
scale. Fortunately, cosmological observational data can provide
the crucial complementary information on absolute neutrino masses,
because massive neutrinos leave imprints on the cosmological
observations, such as the Hubble diagram, CMB temperature power
spectrum and LSS matter power spectrum \cite{NeuRev}.

Within the $\Lambda$CDM model, from Table II one can read $95\%$
upper limit of the total neutrino mass derived from the current
observations, CMB+LSS+SN, $\sum m_{\nu}<0.533~\mathrm{eV}$ ($95\%$
C.L.), which is consistent with the recent results from WMAP5
group \cite{WMAP5GF1,WMAP5GF2}. However, there are degeneracies
between the neutrino mass and other cosmological parameters, such
as the EoS parameters of dark energy \cite{Hannestad} and the
running of spectral index \cite{NeuRun}. Due to the degeneracy
among dark energy parameters and the neutrino mass
\cite{WMAP3GF,Hannestad,XiaMnu}, in the framework of dynamical
dark energy models, the limit on the neutrino mass can be relaxed
to $\sum m_{\nu}<0.974~\mathrm{eV}$ ($95\%$ C.L.) significantly,
as shown in Fig.\ref{fig3}.

It is well known that when neutrinos become non-relativistic at
late time of the universe, they damp the perturbations within
their free streaming scale. Thus the massive neutrinos will
suppress the matter power spectrum at small scale by roughly
$\Delta{P}/P\sim-8\Omega_{\nu}/\Omega_m$ \cite{Suppress}.
Therefore, Lyman-$\alpha$ forest data at small scale can
significantly improve the constraint on the neutrino mass. But
when we use the Ly$\alpha$ data, we should keep in mind on their
unclear systematics right now \cite{WMAP5GF1,FogliWMAP5}. When
including the SDSS Ly$\alpha$ forest power spectrum \cite{Lya}, we
can obtain a much more stringent $2~\sigma$ upper limit $\sum
m_{\nu}<0.161~\mathrm{eV}$ in the $\Lambda$CDM model. %This result
%implies that the degenerate mode of neutrino mass might be
%excluded \cite{FogliWMAP5,SeljakWMAP3}.

Moreover, the Heidelberg-Moscow (HM) experiment, which is
controversial for the time being, uses the half-life of
$0\nu2\beta$ decay to constrain the effective Majorana mass and
this translates to the constraint on the sum of neutrino masses
under some assumptions \cite{HM}, $\sum
m_{\nu}\sim1.8\pm0.6~\mathrm{eV}$ ($95\%$ C.L.). We can find an
obvious tension on the neutrino mass limits from between the
cosmological observations and the HM experiment, which however can
be resolved if the neutrino masses vary during the evolution of
the universe \cite{NeuVary}. In order to be consistent with the
observational data, the neutrino mass must be very small in the
past, but has grown recently in order to agree with the HM
experiment data.

Again, in the RunW model, the limit on the neutrino mass is
relaxed, $\sum m_{\nu}<0.252~\mathrm{eV}$ ($95\%$ C.L.), which
implies the existence of degeneracy between the dark energy
parameters and neutrino mass.

%Results(Others,Inflation)=============================================

\subsection{Inflationary Parameters} \label{Inf}

Inflation, the most attractive paradigm in the very early
universe, has successfully resolved many problems existing in hot
Big Bang cosmology, such as flatness, horizon, monopole problem
and so forth \cite{Guth}. Its quantum fluctuations turn out to be
the primordial density fluctuations which seed the observed large
scale structures and the anisotropies of CMB. Inflation theory has
successfully passed several non-trivial tests. Currently, the
cosmological observational data are in good agreement with a
gaussian, adiabatic and scale-invariant primordial spectrum, which
is consistent with single field slow-roll inflation predictions.

\begin{figure}[t]
\begin{center}
\includegraphics[scale=0.4]{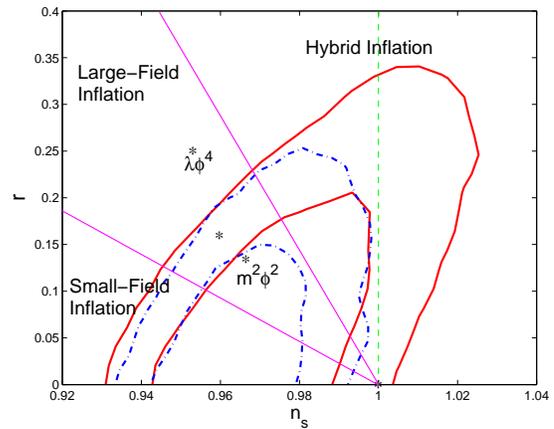}
\caption{$68\%$ and $95\%$ constraints on the panel ($n_s$,$r$)
based on the different Dark Energy models: $\Lambda$CDM model
(blue dash-dot lines) and RunW model (red solid line). The two
magenta solid lines delimit the three classes of inflationary
models, namely, small-field, large-field and hybrid models. The
star points are predicted by HZ spectrum, $m^2\phi^2$ model and
$\lambda\phi^4$ model, respectively. These predictions assume that
the number of e-foldings, $N$, is $50-60$ for $m^2\phi^2$ model
and $64$ for $\lambda\phi^4$ model.\label{fig4}}
\end{center}
\end{figure}

Within the $\Lambda$CDM model, from Fig.\ref{fig3}, we obtain the
limit on the spectral index of $n_s=0.961\pm0.012$ ($1~\sigma$),
which excludes the scale-invariant spectrum, $n_s=1$, and the
spectra with blue tilt, $n_s>1$, at more than $3~\sigma$
confidence level. When considering the gravitational waves, the
latest observational data yield the $95\%$ upper limit of
tensor-to-scalar ratio $r<0.200$. In Fig.\ref{fig4} we show the
two dimensional constraints in ($n_s$,$r$) panel which can be
compared with the prediction of the inflation models. We find that
the Harrison-Zel'dovich-Peebles scale-invariant (HZ) spectrum
($n_s=1$, $r=0$) is still disfavored more than $2~\sigma$
confidence level. And many hybrid inflation models and the
inflation models with ``blue" tilt ($n_s>1$) are also excluded by
the current observations. Furthermore, the single slow-rolling
scalar field with potential $V(\phi)\sim m^{2}\phi^{2}$, which
predicts $(n_s,r)=(1-2/N,8/N)$, is still well within $2~\sigma$
region, while another single slow-rolling scalar field with
potential $V(\phi)\sim \lambda\phi^{4}$, which predicts
$(n_s,r)=(1-3/N,16/N)$, has been excluded more than $2~\sigma$
\cite{WMAP5GF1,KinneyWMAP5,Hiranya}.

However, the tensor fluctuations and the dark energy component,
through the ISW effect, are correlated, which mostly affect the
large scale (low multipoles) temperature power spectrum of CMB
\cite{Xiaplanck,XiaSR}. In the framework of dynamical dark energy
model, we find that the upper limit of $r$ can be relaxed to
$r<0.268$ ($95\%$ C.L.). Furthermore, the $95\%$ confidence level
contour in ($n_s$,$r$) panel will be enlarged consequently and the
distribution extends towards the hybrid inflation region.
Therefore, we can see that the HZ spectrum is consistent with the
latest observational data and many hybrid inflationary models, the
inflationary models with ``blue" tilt ($n_s>1$), which are
excluded in the $\Lambda$CDM model, have revived in the framework
of dynamical dark energy model as illustrated in Fig.\ref{fig4}.

Finally, we explore the constraint on the running of the spectral
index from the latest observational data. When combining the WMAP1
or WMAP3 data with other astronomical data, the previous analysis
have found a significant evidence for large running
\cite{WMAP1Run,WMAP3Run}. Physically this large running would be a
great challenge to the single-field inflation models \cite{KT,BreakSR}.
% described by
%the slow roll expansion which cannot produce this large, negative
%running of scalar spectral index \cite{KT}.
%At that time, two or
%more inflationary history or breaking down the slow roll expansion
%would be needed \cite{BreakSR}.
The combination of the current
observational data yield the limit on the running of the spectral
index of $\alpha_s=-0.019\pm0.017$ ($1~\sigma$) for the
$\Lambda$CDM model and $\alpha_s=-0.023\pm0.019$ ($1~\sigma$) for
the RunW model, respectively. The error is dramatically reduced
compared with the previous results \cite{Xiaplanck,XiaSR},
beneficial from the more accurate observational data. Given the
current data, we find no significant evidence for the large
running of the spectral index.

%Summary===============================================================

\section{Summary}
\label{Sum}

Recently many experimental groups have published their new
observational data, such as temperature and polarization power
spectra of WMAP5 \cite{WMAP5GF1,WMAP5GF2,WMAP5Other}, temperature
power spectrum of ACBAR \cite{ACBAR} and supernovae dataset of
Union compilation \cite{Union}. In this paper we report the
updated constraints on the cosmological parameters from these
latest observational data, such as the EoS of dark energy,
curvature of universe, neutrino mass and inflation parameters.

For dark energy, we explore the constraints on two kinds of
dynamical models. Assuming a flat universe, within the WCDM model,
we find that the latest observational data yield the limit on the
constant EoS of dark energy, $w=-0.977\pm0.056$ ($1~\sigma$). For
RunW model with a flat universe, we find that the best fit model
is $w_0=-1.08$ and $w_1=0.368$ described by the Quintom theory
with EoS across the cosmological constant boundary. And because
the precision of current observations are not good enough to
determine the dark energy EoS conclusively, the dynamical dark
energy models are not excluded and the $\Lambda$CDM model remains
a good fit.

The Quintom scenario, with the particular feature that its EoS can
cross the cosmological constant boundary smoothly, has been
applied to many aspects of cosmology theoretically. Firstly, the
Quintom dark energy models can also give rise to interesting
prediction on the fates of the universe, different from the
Quintessence or Phantom models, such as the cyclic universe
\cite{FengOsc,Oscillating}, an expanding universe with oscillating
EoS. Secondly, applying a Quintom matter for the early universe
can provide a scenario of bouncing cosmology, which can avoid the
notorious issue of initial singularity \cite{Bounce}.

Our results also show that the universe is very close to flatness
and the upper limit on the total neutrino mass is $\sum
m_{\nu}<0.533$ eV ($95\%$ C.L.), from the combination of CMB, BAO
and SN data. Given the efficiency of Ly$\alpha$ forest data on
constraining the total neutrino mass, we also perform a
calculation with the inclusion of the SDSS Lyman-$\alpha$ forest
power spectrum and find that $\sum m_{\nu}<0.161$ eV ($95\%$
C.L.). This result might lead to the exclusion of degenerate
pattern of neutrino mass, when combining the results of neutrino
oscillation experiments. Due to the degeneracy between the
neutrino mass and EoS of dark energy, however in the presence of
dynamics of dark energy, the constraints on $\sum m_{\nu}$ can be
relaxed by a factor of $2$.

Finally, for the inflationary models, within the $\Lambda$CDM
framework, we find that the latest observational data prefer the
inflation models with ``red" tilt, namely, $n_s=0.961\pm0.012$
($1~\sigma$) and small tensor fluctuations, $r<0.200$ ($95\%$
C.L.). Because of the degeneracy between $r$ and EoS of dark
energy, the upper limit on $r$ is relaxed to $r<0.268$ ($95\%$
C.L.) and the parameter space in ($n_s$,$r$) panel are enlarged in
the framework of dynamical dark energy models. Therefore, the
inflationary model with HZ primordial spectrum ($n_s=1$, $r=0$),
some hybrid models and some models with a ``blue" tilt ($n_s>1$),
which are excluded more than $2~\sigma$ confidence level in the
$\Lambda$CDM model, will be consistent with the current
observations now. Furthermore, in our analysis we do not find any
significant evidence for the running of spectrum index.

%Acknowledgments=======================================================

\section*{Acknowledgements}

We acknowledge the use of the Legacy Archive for Microwave
Background Data Analysis (LAMBDA). Support for LAMBDA is provided
by the NASA Office of Space Science. We have performed our
numerical analysis on the Shanghai Supercomputer Center (SSC). We
thank Yi-Fu Cai, Jie Liu and Tao-Tao Qiu for helpful discussions.
This work is supported in part by National Science Foundation of
China under Grant No. 10533010 and 10675136, and the 973 program
No.2007CB815401, and by the Chinese Academy of Science under Grant
No. KJCX3-SYW-N2. HL is supported by the China Postdoctoral
Science Foundation. GZ is supported by National Science and
Engineering Research Council of Canada (NSERC).

%End===================================================================

\end{document}